\documentclass[
preprint,
onecolumn,
amssymb, amsmath
]{revtex4-1}

\newcommand\tcb{}

\usepackage{color}
\usepackage{graphicx}
\usepackage{dcolumn}
\usepackage{bm}

\begin{document}
\preprint{}


\title{Propulsion by a Helical Flagellum in 
\tcb{a Capillary Tube}}


\author{Bin  Liu}
\email{Bin$\underline{\,\,\,\,}$Liu@brown.edu}
\affiliation{School of Engineering, Brown University, Providence, RI 02912, USA}  
\author{Kenneth S. Breuer}
\email{Kenneth$\underline{\,\,\,\,}$Breuer@brown.edu}
\affiliation{School of Engineering, Brown University, Providence, RI 02912, USA}  
\author{{Thomas} R. Powers}
\email{Thomas$\underline{\,\,\,\,}$Powers@brown.edu}\affiliation{School of Engineering, Brown University, Providence, RI 02912, USA}
\affiliation{Department of Physics, Brown University, Providence, RI 02912, USA}

\date{\today}
\begin{abstract}
We study the microscale propulsion of a rotating helical filament confined by a cylindrical tube, using a boundary-element method for Stokes flow that accounts for helical symmetry. We determine the effect of confinement on swimming speed and power consumption. Except for a small range of tube radii at the tightest confinements, the swimming speed at fixed rotation rate increases monotonically as the confinement becomes tighter. At fixed torque, the swimming speed and power consumption depend only on the geometry of the filament centerline, except at the smallest pitch angles for which the filament thickness plays a role.
We find that the `normal' geometry of \textit{Escherichia coli} flagella is optimized for swimming efficiency, independent of the degree of confinement. The efficiency peaks when the arc length of the helix within a pitch matches the circumference of the cylindrical wall. We also show that a swimming helix in a tube induces a net flow of fluid along the tube.
\end{abstract}

\pacs{}
\keywords{}
\maketitle

Bacteria commonly swim in porous materials such as soil, mucus, and tissue. A better understanding of the fluid mechanics of bacterial swimming motility in porous media could have implications for a host of environmental and medical problems. For example, bacteria are used to consume pollutants in contaminated aquifers, and it has been shown that motile  bacteria are less likely to adhere to the soil particles when there is modest groundwater flow~\cite{CamesanoLogan1998}. Likewise, the mucin polymers in the mucus layers lining the stomach wall seem to promote swimming behavior in the outer layer of mucus, which in turn prevents bacteria from aggregating in the layer of mucus immediately adjacent to the stomach wall~\cite{Caldaraetal2012}.  Finally, motile spirochetes are much more likely to penetrate tissues and proliferate when compared to non-motile spirochetes~\cite{Lux01102001}. These examples have motivated experimental and theoretical studies of swimming motility in confined geometries. Early measurements investigated the change in swimming speed for \textit{Escherichia coli} bacteria moving parallel or perpendicular to nearby flat glass surfaces~\cite{FrymierFord1997}. Studies of swimming in microfabricated channels showed that swimming speed of bacteria swimming parallel to the channels was unaffected for channel depths greater that 10~\,$\mu$m, and increased by about 10\% for channel depths of 3~\,$\mu$m, which is comparable to the size of the cell~\cite{BiondiQuinnGoldfine1998}. These observations are in qualitative accord with boundary-element computations that show a modest increase in swimming speed for rotating helices pushing a cell body near plane boundaries~\cite{ramia_etal1993}.
Other experimental studies of \textit{E. coli} in capillary tubes have shown that the swimming becomes unidirectional when the tube diameter is comparable to the cell size, as well as multi-cell behavior such as aggregation and swimming clusters~\cite{Liu:1997}.
More recently, channels in the surface of porous ceramic covered with a film of water have been used as controlled model systems to study motility of bacteria in soil~\cite{Dechesne10082010}, the role of the nature of material properties of the surfaces has been examined~\cite{diluzio05}, and it has been shown that \textit{E. coli} and \textit{Bacillus subtilis} can swim in microfluidic channels with widths as small as 30\% of the cell diameter~\cite{Maennik01092009}. 

\begin{figure}[h]
\begin{center}
\includegraphics[width=0.6\textwidth]{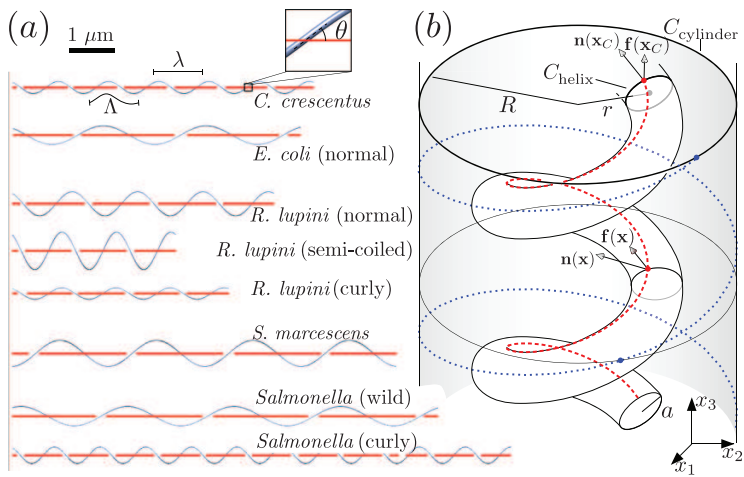}
\caption{ (Color online.) (a) Typical bacterial flagella of different helical \tcb{geometries}
and handedness~\cite{Koyasu:1984, Darnton:2004, Scharf:2002, Darnton:2007, Hesse:2009, Macnab:1977}. (b) Model helical swimmer confined by \tcb{a} cylind\tcb{er}. Physical quantities, such as force density $\mathbf{f}(\mathbf{x})$ or the surface normal $\mathbf{n}(\mathbf{x})$,  are identical up to a rotation along the dashed helix on the filament surface or along the dotted helix on the tube surface  due to the helical symmetry.
Therefore, knowledge of physical quantities on the circular contours $C_\textrm{helix}$ and $C_\textrm{cylinder}$ is sufficient to reconstruct the values of these quantities over the entire surface of the helix and the cylinder. 
\label{fig:diagram}}
\end{center}
\end{figure}

In this letter we consider the simplest model system for studying the interplay between motility and porosity by calculating the swimming speed of an infinite rotating helical flagellum~\cite{GH:1955, Lighthill:1976, Purcell:1977, LP:2009} in a cylindrical tube at zero Reynolds number. Since there are a variety of helical geometries that commonly occur in bacterial species~\cite{Koyasu:1984, Darnton:2004, Scharf:2002, Darnton:2007, Hesse:2009, Macnab:1977} (Fig.~\ref{fig:diagram}a), we study how the speed of a rotating helix in a tube depends on the pitch angle of the helix. Bacterial flagella rotate as rigid bodies, but as a comparison we also study propagating helical bending waves with fixed wave speed. Because many bacteria in Nature swim at constant torque~\cite{Manson:1980}, we also calculate swimming speed for a rotating rigid helix subject to a fixed torque. We examine how efficiency depends on confinement and helix geometry. Finally, for one helix geometry, we compute the flow field to high accuracy and show there is net flux induced by the motion of the swimmer. 
Our work is complementary to a recent numerical study of the motility of a spherical squirmer in a capillary tube~\cite{ZhuLaugaBrandt2013}. Since our boundary element method exploits helical symmetry to speed up computation~\cite{LBP:2013}, we enforce helical symmetry by assuming the helix is infinitely long and supposing the centerline of the helix coincides with the axis of the tube. For finite-length swimmers near the wall of a cylinder, we expect curved trajectories since swimmers near a flat wall swim in circles~\cite{lauga06}, but that problem is beyond the scope of this letter.

\tcb{Figure~\ref{fig:diagram}(b) displays the geometry of our model. A filament of radius $a$ rotates about the $x_3$ axis with rate $\Omega$ and translates along the $x_3$-direction with swimming speed $V$, which is determined by the condition that the axial force on the filament vanishes~\cite{LPB:PNAS2011}. The helical centerline of the filament has radius $r$ and pitch $\lambda$. The arc length $\Lambda$ of one helical pitch is $\Lambda=\sqrt{4\pi^2 r^2 + \lambda^2}$, and the pitch angle $\theta$ is given by $\cos\theta=\lambda/\Lambda$ (Fig.~\ref{fig:diagram}a). The axis of symmetry of the helical centerline is coaxial with the confining cylinder, which has radius $R$. In our calculations we use aspect ratios $a/\Lambda\approx0.01$, which is approximately ten times larger than the aspect ratio for an \textit{E. coli} flagellar filament, but comparable to the aspect ratio for a bundle of filaments~\cite{Koyasu:1984, Darnton:2004, Scharf:2002, Darnton:2007, Hesse:2009, Macnab:1977}. } 

We use a boundary-element method (BEM)~\cite{Pozrikidis:1992td} to solve for the swimming speed. BEMs for Stokes flow reduce a 3d problem to a 2d problem since they amount to determining the distribution of singular solutions on the 2d boundaries of the system, which in our case are the surface of the helical filament and the surface of the tube. We can further reduce the problem to a 1d problem by exploiting helical symmetry~\cite{LBP:2013}. The helical symmetry means that there are helical curves on the surface of the filament [dashed line, Fig.~\ref{fig:diagram}(b)] and the cylinder [dotted line, Fig.~\ref{fig:diagram}(b)] along which scalar quantities such as pressure are constant and vector quantities such as force per unit length rotate at a fixed rate with $x_3$.  Thus, helical symmetry allows us to reduce the problem to determining effective force densities along the circular contours $C_\textrm{helix}$ and $C_\textrm{cylinder}$. Since there are interactions between parts of the filament (and wall) at different values of $x_3$, we must integrate over the $x_3$ direction to obtain the effective force densities on the contours $C_\textrm{helix}$ and $C_\textrm{cylinder}$. Thus, we cut off the filament with a finite number of pitches $\kappa$, where $\kappa=40$ is taken large enough that finite-size effects are negligibly small~\cite{LBP:2013}. Once the force densities on the filament and tube are known, the full flow field may be calculated. The reduction of order by helical symmetry substantially reduces the computational time for a given resolution.

\begin{figure}[h]
\begin{center}
\includegraphics[width=0.8\textwidth]{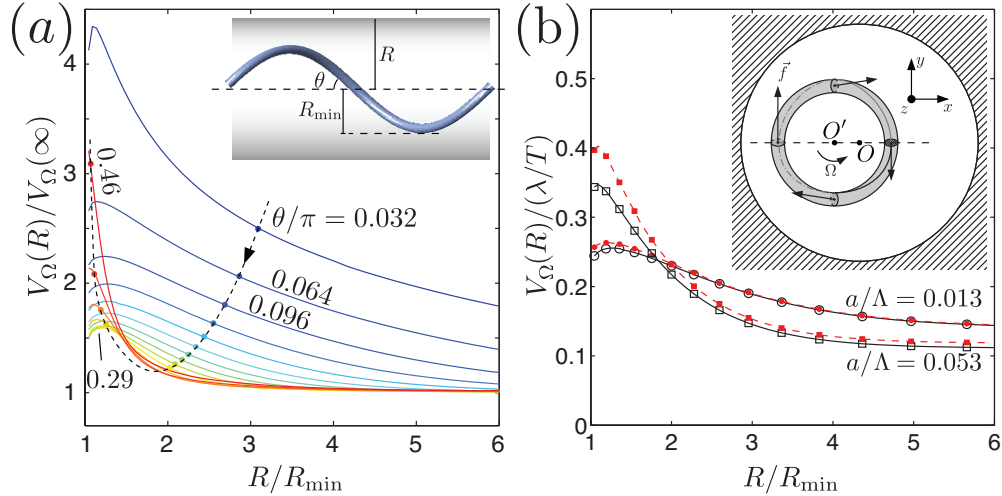}
\caption{(Color online.) Swimming speeds for helical swimmers in a cylindrical tube with radius $R$. (a) Dimensionless swimming speed versus dimensionless tube radius for rigid-body rotation, for various pitch angles $\theta$ and  aspect ratio $a/\Lambda=0.013$. The curves are equally spaced in $\theta/\pi$: as you trace along the dashed line in the direction of the arrow,  $\theta/\pi$ increases by 0.032 for each successive curve. 
(b) Swimming speed \tcb{normalized by no-slip translation speed (see text)} for \tcb{a particular} fixed amplitude $\theta=0.16\pi$ for rigid-body rotation (solid curves) and helical waves (dashed curves). Results are shown for two different aspect ratios, $a/\Lambda=0.013$ (circles) and $a/\Lambda=0.053$ (squares). The inset shows the force distribution in the more general situation when the helix is off-center. The dashed line is a symmetry axis. 
}\label{fig:fixrot}
\end{center}
\end{figure}

Let $V_\Omega(R)$ denote the swimming speed of a helical filament with rotational speed $\Omega$ in a tube of radius $R$. Due to the negligible fluid inertia, the speed $V_\Omega(R)$ is linearly proportional to $\Omega$. The minimum radius of the tube is $R_\textrm{min}=r+a$, the radius for which the filament just touches the tube wall. Figure~\ref{fig:fixrot}(a) shows the swimming speed at fixed rotation rate normalized by the unconfined swimming speed at same rotation rate, $V_\Omega(\infty)$. For fixed rotation rate, confinement always increases the swimming speed relative the the unconfined case.
For most values of $R/R_\textrm{min}$,  the ratio $V_\Omega(R)/V_\Omega(\infty)$ increases monotonically with decreasing $R/R_\textrm{min}$. There is a slight decrease in $V_\Omega(R)/V_\Omega(\infty)$ as $R/R_\textrm{min}$ approaches 1. The numerical results at this limit are ensured to converge robustly at a third order \cite{LBP:2013} by refining the spatial grids to be much smaller than the gap between the filament and the tube wall. It is natural to expect that the no-slip speed of a corkscrew, $V_\textrm{no-slip}=\lambda\Omega/(2\pi)$, is an upper bound for the swimming speed. But it is not a very tight upper bound: even at the maximum speed, which occurs when $R/R_\textrm{min}\approx1$, the swimming speed is only a small fraction of the no-slip speed, $V_\Omega(R)\approx0.1V_\textrm{no-slip}$. The relative enhancement due to confinement is greatest at the smallest and the largest pitch angles, with the effect of confinement evident over a larger range of tube radii for the smaller pitch angles. 

The filament  diameter affects swimming speed for unconfined swimming (see e.g.~\cite{LBP:2013}). Swimming speed decreases with thickness in the absence of a confining tube. But for small enough radius, a confining tube can make thick filaments swim \textit{faster} than thin filaments. Since this effect is best displayed by plotting the confined swimming velocities relative to a standard that is independent of filament radius $a$, the swimming velocities in Fig.~\ref{fig:fixrot}(b) are normalized by the no-slip velocity. We see that the thicker filaments with $a/\Lambda=0.053$ swimmer faster than the filaments with $a/\Lambda=0.013$ once the tube radius is less than about twice its minimum. 

The effect of filament diameter on propulsion in a tube can also be illustrated by comparing rigid-body rotation with a swimming stroke that is a propagating helical wave. In a propagating helical wave with frequency $\Omega$, the filament deforms such that its centerline coincides with the centerline of a rigid helical filament rotating with speed $\Omega$, but the cross-sections of the filament do not rotate as the wave advances. Since the centerline motion is the same for rigid-body motion and the helical wave, and since we demand the shape of the cross-section of the filament transmitting the helical wave be the same as the shape in rigid-body rotation, differences in swimming speed between the two strokes arise from finite-thickness effects. To calculate the swimming speed, we need to specify the velocity on the surface of the filament. For either swimming stroke, we define a  frame $\{x_1,x_2\}$ with origin at the point the helical centerline pierces the cross section.  Note that although the origin of this frame moves in a circle as the rigid filament rotates or the deformable filament bends, the basis vectors do not rotate. Let $\{\rho, \alpha\}$ denote the polar coordinates of this frame.  For rigid-body motion, the surface velocity is $\mathbf{v}_\mathrm{rb}=\Omega\hat{\mathbf{x}}_3\times[\mathbf{r}+\rho(\alpha,t,x_3)\hat{\bm\rho}]$, where $\mathbf{r}=\mathbf{r}(t,x_3)$ denotes the position of the helical centerline, and $\rho(\alpha,t,x_3)$ is the radius vector  from the origin to the point of interest with angle $\alpha$ on the surface of the filament at cross section $x_3$ at time $t$. To determine the surface velocity field for a filament subject to helical waves,  we demand that each cross section deforms such that there is no rotation of material lines in the cross section, and such that the \textit{shape} of the cross-section rotates like the cross-section of the rigidly rotating helix: $\rho(\alpha+\Omega\mathrm{d}t,t+\mathrm{d}t)=\rho(\alpha,t)$, or $\partial\rho/\partial t|_\alpha=-\Omega\partial\rho/\partial\alpha|_t$. Thus, the surface velocity for the helical wave is
$\mathbf{v}_\textrm{hw}=\Omega\left(\hat{\mathbf{x}}_3\times\mathbf{r}-{\partial\rho}/{\partial\alpha}\hat{\bm\rho}\right)$.
 
Figure~\ref{fig:fixrot}(b) shows the comparison between swimming speeds for rigid-body rotation and a helical wave for two different aspect ratios. As expected, there is little difference between the swimming speeds for the two strokes for the thin filaments with $a/\Lambda=0.013$ (circles). However, for the thick filaments with $a/\Lambda=0.053$, there is a small but non-negligible difference between the surface velocities for the two strokes, leading to a noticeable difference in swimming speeds, which is greatest when the confinement is greatest. 

The helix is neutrally stable against small displacements away from the center of the tube. To see why, consider the offset $OO'$  along the (arbitrary) $x$-axis (fig.~\ref{fig:fixrot}(b), inset). Rotating the entire system by $\pi$ about the axis $OO'$ and reversing the direction of rotation is a symmetry. Using this symmetry we may deduce that if we anchor the helix such that its central axis coincides with $O'$,  the hydrodynamic forces per length obey $(f_x(x,y),f_y(x,y))=(-f_x(x,-y),f_y(x,-y))$. Thus the net $x$-component of the hydrodynamic force on a period of the helix vanishes: the hydrodynamic forces do not push the helix away or toward the center of the tube. But since the $y$ component of the force on a period does not vanish, the helix will trace out a helical trajectory if the anchoring is released.

\begin{figure}[h]
\begin{center}
\includegraphics[width=0.8\textwidth]{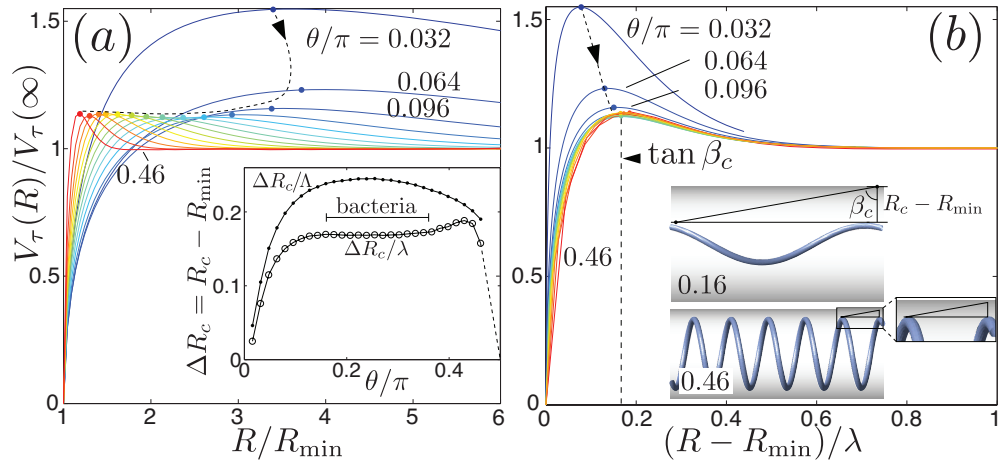}
\caption{(Color online.) Swimming speed of rotating helical swimmers with fixed torque $V_\tau(R)$. (a) The solid curves are equally spaced in $\theta/\pi$,
with the dashed line showing the direction of increasing $\theta$. The speed $V_\tau(R)$ peaks at $R=R_c$. The dependence of $R_c$ on $\theta$ is shown in the inset, normalized by arc length $\Lambda$ (`\textbullet') and wavelength $\lambda$ (`$\circ$'). The scale bar represents the range of typical $\theta$'s for bacteria. (b) Re-plot of (a) with $R$ normalized by $\lambda$; again the curves are equally spaced in $\theta/\pi$. When $\theta$ is sufficiently large ($\theta \gtrsim 0.1\pi$), the data collapse. 
The inset shows 
two different helical geometries ($\theta=0.16\pi$ and $0.45\pi$) with the same values of $\beta$.
\label{fig:fixtorq}}
\end{center}
\end{figure}

We now turn to the effect on confinement on the swimming speed $V_\tau$ of rigid helices rotated with fixed torque $\tau$ (Fig.~\ref{fig:fixtorq}). When the torque is held fixed, the behavior is very different from the case with fixed rotation velocity, especially in the limit of greatest confinement. When $R$ gets close to $R_\textrm{min}$,
the swimming speed $V_\tau$ decreases as $R$ decreases, and  
eventually vanishes when the filament touches the wall when $R=R_\textrm{min}$. Due to the no-slip boundary condition, an infinite torque is required to turn the helix when it just touches the tube wall. A similar effect was shown previously by Felderhof for axisymmetric swimmers in a cylindrical tube~\cite{Felderhof:2010}.

We also find that there is a critical radius $R=R_c$ such that the speed $V_\tau$ reaches a peak before relaxing back to the unconfined limit at large $R$. The relative enhancement due to confinement is weakly dependent on pitch angle except for small pitch angles, $\theta\lesssim0.1\pi$.
The inset  of Fig.~\ref{fig:fixtorq}(a) shows how the critical gap, $\Delta R_c=R_c-R_\textrm{min}$, depends on pitch angle $\theta$ for fixed contour period $\Lambda$ (dots) or fixed pitch $\lambda$ (circle).
For fixed $\Lambda$, the gap $\Delta R_c$ has a peak for a helix of moderate pitch, $\theta= \pi/4$. For fixed $\lambda$, the gap is almost invariant for a wide range of $\theta$, especially for the range that corresponds to the commonly observed geometrical parameters of bacteria, indicated by a horizontal bar. This result suggests that the wavelength $\lambda$ is an important parameter governing the effect of confinement on helical swimming at fixed torque. Therefore,  we plot $V_\tau(R)/V_\tau(\infty)$ as a function of $(R-R_\mathrm{min})/\lambda$ in Fig.~\ref{fig:fixtorq}(b). This plot reveals that the swimming speeds partially collapse onto a single curve, with the best collapse resulting when the pitch angle is modest, $\theta\gtrsim0.1\pi$. 
Note that the fact that $\lambda$ enters the scaling variable implies that the hydrodynamic interactions between successive turns of the helix are important along with the hydrodynamic interactions that arise from the confining tube.
The controlling geometrical factor is conveniently visualized as the angle $\beta=\tan^{-1}[(R-R_\textrm{min})/\lambda]$. The inset of Fig.~\ref{fig:fixtorq}(b) shows two swimmers of different geometry  ($\theta=0.16\pi$ and $\theta=0.45\pi$) but the same value of $\beta=\beta_c$, the angle at which there is  peak enhancement due to confinement. For a fixed length of contour $\Lambda$ in one pitch, the helical radius $r$ is comparable to the filament radius $a$ when the pitch angle is small $\theta\lesssim0.1/\pi$. In this limit we expect finite-thickness effects to govern the swimming speed and we do not expect the speed to be determined solely by the geometry of the helical centerline. This expectation is consistent with the fact that the curves in Fig.~\ref{fig:fixtorq}(b) do not collapse for small $\theta$. To be more general, the regime in which the filament thickness is important has $a/R\gtrsim0.3$ for $a/\Lambda$ ranging from $0.013$ to $0.053$. This criterion also coincides with the condition that the radius of the hollow center of the helix $R-a$ be comparable to the filament diameter $2a$.

\begin{figure}[h]
\begin{center}
\includegraphics[width=0.6\textwidth]{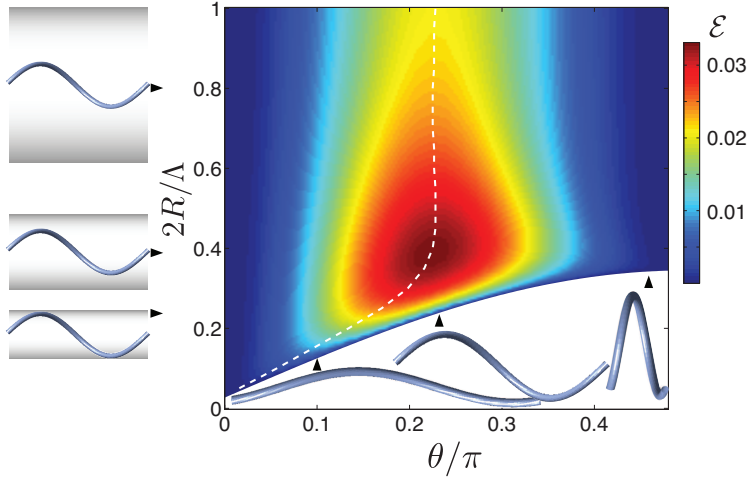}
\caption{(Color online.) Dependence of efficiency $\mathcal{E}$ on the pitch angle $\theta$ and dimensionless tube diameter $R/\Lambda$. The peak of $\mathcal{E}$ at given $R/\Lambda$ is shown in the dashed curve. 
The bottom boundary of the colored region corresponds to the minimum value of the tube radius for a given pitch angle and fixed $\Lambda$.
}\label{fig:fixpwr}
\end{center}
\end{figure}

Confinement has a strong effect on the efficiency of helical swimming. We define efficiency as follows. For a given geometry and tube radius, we prescribe the rotation speed $\Omega$ and compute the swimming speed $V_\Omega$ and the torque $\tau$ required to rotate the helix at speed $\Omega$. Then we compute the power $P=\Omega\tau$. Since we wish to compare the power required to propel helices of different geometries, we compare $P$ with the power $P_\textrm{min}$ required to drag a straight cylinder along the center of the tube at velocity $V_\Omega$. This power is also computed numerically using our BEM scheme. The efficiency is thus $\mathcal{E}\equiv P_\textrm{min}/P$.
As shown in Fig.~\ref{fig:fixpwr}, the efficiency $\mathcal{E}$ peaks for a normal flagellum ($\theta\approx \pi/4$). Although the value of the efficiency depends strongly on the degree of confinement, the pitch angle for which the efficiency is maximized is
almost independent of the tube radius $R$.
Note also that the efficiency is 
maximized when the diameter of the confining tube is roughly half the helix contour length, $2R/\Lambda \equiv 0.4$. It should be noted that this optimized tube size also coincides with the maximal helix diameter $d_\textrm{max}$ at given $\Lambda$:  $2R\sim d_\textrm{max}=\Lambda/\pi$. 

\begin{figure}[h]
\begin{center}
\includegraphics[width=0.6\textwidth]{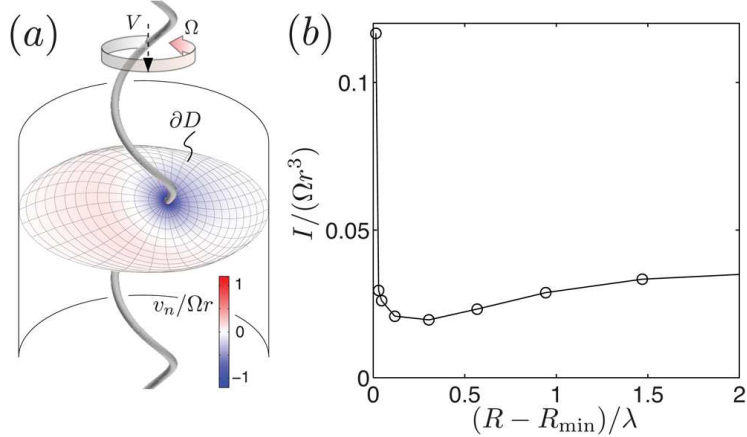}
\caption{(Color online.) Fluid transport due to a swimming helix  with fixed rotation rate $\Omega$.
(a) The flux is computed across a curved surface that smoothly connects the circles $C_\mathrm{helix}$ and $C_\mathrm{cylinder}$. The flow component $v_n$ normal to the surface is shown by the greyscale (color online).
(b) A net flux in the opposite direction of swimming is induced. Here  $\theta=0.75$ \tcb{and} $a/\Lambda=0.013$}\label{fig:flux}
\end{center}
\end{figure}

Finally we turn to the question of whether fluid is entrained along with the swimmer, or forced past it in the direction opposite to swimming. When a sphere is towed in an infinite fluid, the induced flow causes tracer particles to experience a net displacement opposite to the direction of motion of the sphere~\cite{Darwin1953}. In the case of model swimmers, it has recently been shown that the net displacement of a tracer particle depends on the initial distance between the swimmer and the tracer particle~\cite{Pushkin2013}. 
To determine the  direction of fluid transport in our problem, we use the helical symmetry and our high spatial resolution to calculate the flow induced by the swimmer. We calculate the flux across a surface $\partial D$ which is compatible with the helical symmetry. The surface smoothly extends from a contour circulating the filament cross-section, $C_\textrm{helix}$ [Fig.~\ref{fig:diagram}(b)], to \tcb{a contour on} the tube wall,   $C_\textrm{cylinder}$ [Fig.~\ref{fig:diagram}(b)]. The flux is given by
$I=\int_{\partial D} \textrm{d}s \mathbf{v}(\mathbf{r})\cdot \mathbf{n}=\int_{\partial D} \textrm{d}s v_n(\mathbf{r})$, 
where $\mathbf{n}$ is the surface norm, and $v_n=\mathbf{v} \cdot \mathbf{n}$ is the flow component normal to that surface.
The surface is chosen such that no interpolation or extrapolation among the mesh points is necessary for evaluating the flux numerically.

The dependence of the flux $I$ on the confinement is shown in Fig.~\ref{fig:flux}(b). Even though the helix swims free of external force, there is net flux produced in the direction opposite to the swimming direction. The flux does not seem to vanish as it saturates as the tube size $R$ increases. Interestingly, at a given rotation rate $\Omega$, there exists a minimum flux with varying tube size, which is reminiscent of the critical tube dimension associated with the peak enhancement in swimming with fixed torque. 

To conclude, we have used a boundary-element method to calculate the swimming speed of a helical filament at fixed rotation speed or fixed torque. Our results are in qualitative accord with previous calculations of swimming near boundaries, in that we also find that swimming speed is enhanced very near a wall for prescribed rotation~\cite{ramia_etal1993}, and vanishes for prescribed torque~\cite{Felderhof:2010}. Likewise, our calculations are qualitatively consistent with experiments that show a modest in swimming speed of confined bacteria~\cite{BiondiQuinnGoldfine1998}, and our results agree with Liu \textit{et al.} within their experimental error~\cite{Liu:1997}. 
Although our results show that the pitch angle at which maximum efficiency is achieved does not depend strongly on confinement, and is close to that of the normal polymorphic form of the \textit{E. coli} flagellar filament. We have also show that swimming helical filaments drive a flux of fluid in the opposite direction. 

There are several direction in which our work can be extended. The dependence of flux on helical geometry and tube size should be more thoroughly explored. It would be interesting to remove the helical symmetry and study the trajectory of helical swimmers along curved walls, and to include the effects of the cell body. Applications to bioremediation could be explored by adding an ambient flow and studying the effect of confinement and swimming on cell adhesion to walls. Finally, there are different geometries which must be governed by the same underlying physics, such as the swimming of a spermatozoa near a wall. This problem is often modeled by an infinite swimming sheet near a wall~\cite{Katz:1974}, and recently the effect of fluid viscoelasticity on the efficiency of swimming near a wall has been examined in numerical computations~\cite{Chrispell:2013}. It would be interesting to study the effects of viscoelasticity on swimming helices confined to capillary tubes.

We acknowledge helpful discussions with M. A. Dias and H. C. Fu. This work was supported by the National Science Foundation Grant No. CBET-0854108.

\end{document}